\documentstyle [fleqn,epsf] {article}
\input epsf
\def\gb{\overline g}
\def\dtau{\Delta\tau}
\def\si{\sigma}
\def\ep{\epsilon}
\def\lep{\left(}
\def\rip{\right)}
\def\up{\uparrow}
\def\dwn{\downarrow}
\def\be{\beta}
\def\De{\Delta}
\def\al{\alpha}
\def\frac#1#2{{#1 \over #2}}
\def\de{\delta}
\def\V{\cal V}
\begin{document}
\noindent
{\Large\bf Simulation of the electron-phonon interaction in
infinite~dimensions}

\vspace{0.2cm}

\noindent J.~K.~Freericks

\vspace{0.2cm}

\noindent Department~of~Physics,~University~of~California,~Davis,~CA~95616

\vspace{0.2cm}

\noindent M.~Jarrell

\vspace{0.2cm}

\noindent Department~of~Physics,~University~of~Cincinnati,~Cincinnati,~OH~45221
\vspace{1.5cm}

\noindent{\bf Abstract.} The electron-phonon interaction corresponding to
the Holstein model (with Coulomb repulsion)
is simulated in infinite dimensions using a novel quantum
Monte Carlo algorithm.  The thermodynamic phase diagram includes
commensurate charge-density-wave phases, incommensurate charge-density-wave
phases, and superconductivity.  The crossover from a weak-coupling
picture (where pairs both form and condense at $T_c$) to a strong-coupling
picture (where preformed pairs condense at $T_c$) is illustrated with
the onset of a double-well structure in the effective phonon potential.

\bigskip
\noindent{\bf 1. Infinite-dimensional formalism}
\bigskip

\noindent Strong electron-electron correlations are responsible for many
important and exotic phenomena in condensed-matter systems including
superconductivity, magnetism, heavy fermions, {\it etc}.  Strongly correlated
electronic systems are those in which the average electronic correlation energy
is equal to or larger than the electronic kinetic energy.  Exotic phenomena
arise from the competition of simultaneously minimizing the kinetic and
potential energy of the electrons.  Models of these systems usually do not
have analytic solutions.  However, recently, Metzner and
Vollhardt~\cite{metzner_vollhardt} discovered that these many-body problems
simplify in the limit of infinite spatial dimensions.  The limit must be
taken in such a fashion that the electronic kinetic energy remains finite,
so that the effects of the strong electron correlations remain.

Consider the electronic kinetic energy determined by a tight-binding model
with hopping between nearest-neighbor sites (with hopping integral $t$) on
a hypercubic lattice in $d$ dimensions.  The band structure $\epsilon({\bf k})$
becomes
\begin{equation}
\epsilon({\bf k})=-2t\sum_{i=1}^{d} \cos {\bf k}_i\quad\quad .
\label{eq: band}
\end{equation}
In the infinite-dimensional limit ($d\rightarrow\infty$) the set of
$\{\cos{\bf k}_i\}$ can be thought of as ``random'' numbers distributed
between $-1$ and $1$ for a general point in the $d$-dimensional Brillioun
zone.  The sum of $d$ ``random'' numbers grows as $\sqrt{d}$, so the band
structure remains finite if the hopping integral scales as
$t=t^*/2\sqrt{d}$~\cite{metzner_vollhardt}.  Furthermore
the central limit theorem states that the density of states corresponding
to this band structure [$\rho(y)$] becomes a Gaussian distribution
\begin{equation}
\rho(y)={1\over\sqrt{\pi}t^*}\exp \Big ( -{y^2\over t^{*2}} \Big )\quad ,
\label{eq: gaussian}
\end{equation}
in the infinite-dimensional limit.  The number of nearest neighbors ($2d$)
diverges, but the hopping between nearest neighbors $(t=t^*/2\sqrt{d})$
vanishes in such a fashion to maintain a finite kinetic energy for the
electrons.

The phonon density of states has a very different behavior in the
infinite-dimensional limit.  The phonon density of states for the Debye model
(phonons with a linear dispersion from zero frequency to $\omega_D$) is
\begin{equation}
N(\omega)={C_d\over \omega_D} \Big [ {\omega\over \omega_D} \Big ]^{d-1}\quad ,
\quad 0\le \omega\le\omega_D\quad ,
\label{eq: debye}
\end{equation}
in $d$ dimensions.  In the limit as $d\rightarrow\infty$, the phonon
density of states becomes a delta function at the Debye frequency.

These two observations for the electron and phonon densities of states
motivate one to examine the Holstein-Hubbard model~\cite{holstein,hubbard}
(in which the electrons couple to localized phonons)
as the simplest electron-phonon model in infinite-dimensions:
\begin{eqnarray}
  H &=& - {{t^*}\over{2\sqrt{d}}} \sum_{\langle j,k\rangle \sigma} (
   c_{j\sigma}^{\dag }
  c_{k\sigma} + c_{k\sigma}^{\dag  } c_{j\sigma} ) + \sum_j  (gx_j-\mu)
  (n_{j\uparrow}+n_{j\downarrow}-1) \cr
&+& U_c\sum_j(n_{j\uparrow} -{1\over 2})
   (n_{j\downarrow}-{1\over 2}) + {{1}\over{2}} M \Omega^2 \sum_j x_{j}^2 +
  {{1}\over{2}} \sum_j {p_{j}^2\over M}\quad .
\label{eq: holhubham}
\end{eqnarray}
Here $c_{j\sigma}^{\dag}$ ($c_{j\sigma}$) creates (destroys) an electron at
site $j$ with spin $\sigma$, $n_{j\sigma}=c_{j\sigma}^{\dag}c_{j\sigma}$ is
the electron number operator, and $x_j$ ($p_j$) is the phonon coordinate
(momentum) at site $j$.  The hopping matrix elements connect the nearest
neighbors of a hypercubic lattice in $d$-dimensions and the
unit of energy is chosen to be this rescaled matrix
element $t^*$.  The phonon has a mass $M$ (chosen to be $M=1$), a
frequency $\Omega$, and a spring constant $\kappa\equiv M\Omega^2$ associated
with it.  The electron-phonon coupling constant (deformation potential)
is denoted by $g$ so that the effective electron-electron attraction
becomes the bipolaron binding energy
\begin{equation}
  U\equiv - {{g^2}\over{M\Omega^2}}=-{g^2\over\kappa} \quad .
 \label{eq: udef}
\end{equation}
The Coulomb repulsion is represented by a local Hubbard interaction $U_c$ and
the chemical potential is denoted by $\mu$ with particle-hole symmetry
occurring
for $\mu=0$.

The observation of Metzner and Vollhardt~\cite{metzner_vollhardt} is that
the many-body problem also simplifies in the infinite-dimensional limit---both
the self energy and the irreducible vertex functions become independent of
momentum, and are functionals of the interacting Green's
function~\cite{metzner_vollhardt,schweitzer_czycholl,metzner}.  The Green's
function, self energy, and irreducible vertices still retain their complicated
time (frequency) dependence.

The many-body problem is solved by mapping it onto an auxiliary impurity
problem~\cite{brandt_mielsch,okhawa} in a time-dependent field
that mimics the hopping of an electron onto
a site at time $\tau$ and off the site at a time $\tau '$.  The
action for the impurity problem is found by integrating out all of the
degrees of freedom of the other lattice sites in a path-integral
formalism~\cite{georges_kotliar}.  The lattice is viewed as a reservoir of
electrons that can hop onto and off of the local site.  Once an electron
hops off of the local site, it never returns, because the number of paths
that loop through the local site are a factor of $1/d$ smaller than the number
of paths that do not loop through the local site.  Therefore, the effective
action for the impurity problem becomes
\begin{eqnarray}
S &=& \sum_{\sigma} \int_0^{\beta} d\tau \int_0^{\beta} d\tau '
c_{\sigma}^{\dag}(\tau)G_0^{-1}(\tau-\tau ')c_{\sigma}(\tau ')\cr
&+&\sum_{\sigma}\int_0^{\beta}d\tau[gx(\tau)-\mu][n_{\sigma}(\tau)-{1\over 2}]
\cr
&+& U_c\int_0^{\beta}d\tau [n_{\uparrow}(\tau)-{1\over 2}]
[n_{\downarrow}(\tau)-{1\over 2}] + {M\over2}\int_0^{\beta}d\tau
[\Omega^2x^2(\tau)+\dot x^2(\tau)]
\label{eq: seff}
\end{eqnarray}
where $G_0^{-1}$ is the ``bare'' Green's function that contains
{\it all of the dynamical information of the other sites of the lattice}.
The interacting Green's function, defined to be
\begin{equation}
G(i\omega_n)\equiv \int_0^{\beta} d\tau e^{i\omega_n\tau}G(\tau)\quad , \quad
G(\tau)=-{{\rm Tr} \langle e^{-\beta H}T_{\tau} c(\tau)
c^{\dag}(0)\rangle\over {\rm Tr} \langle e^{-\beta H}\rangle } ~,
\label{eq: greendef}
\end{equation}
then satisfies Dyson's equation
\begin{equation}
G_n^{-1}\equiv G^{-1}(i\omega_n) = G_0^{-1}(i\omega_n)-\Sigma (i\omega_n).
\label{eq: gdef}
\end{equation}

A self-consistency relation is required in order to determine the bare
Green's function $G_0$.  This is achieved by mapping the impurity problem
onto the infinite-dimensional lattice thereby equating the full Green's
function for the impurity problem with the local Green's function for
the lattice
\begin{eqnarray}
G_{jj}(i\omega_n)&=&\sum_{\bf k} G({\bf k},i\omega_n) = \sum_{\bf k}
[i\omega_n+\mu-\epsilon({\bf k})-\Sigma(i\omega_n)]^{-1}\cr
&=& F_{\infty}[i\omega_n+\mu -\Sigma(i\omega_n)].
\label{eq: gloc}
\end{eqnarray}
Here $F_{\infty}(z)$ is the scaled complimentary error function of a complex
argument.
\begin{eqnarray}
F_{\infty}(z)&\equiv&{1\over{\sqrt{\pi}}}\int_{-\infty}^{\infty} dy
{\exp(-y^2)\over{z-y}} \cr
&=&-i{\rm sgn}[{\rm Im}(z)]\sqrt{\pi}e^{-z^2}{\rm erfc}
\{-i{\rm sgn}[{\rm Im}(z)]z\}.
\label{eq: fdef}
\end{eqnarray}
The dynamics of the (local) impurity problem is identical to the dynamics
of the Anderson impurity
model~\cite{schweitzer_czycholl,brandt_mielsch,okhawa,georges_kotliar,jarrell}.
This many-body problem can be solved exactly with the quantum Monte Carlo
(QMC) algorithm
of Hirsch and Fye~\cite{hirsch_fye} (see the next section).  The impurity is
self-consistently embedded in the host, since
it must satisfy the self-consistency relation in Eq.~(\ref{eq: gloc}).
Note that this mapping of the infinite-dimensional lattice problem onto a
single-site impurity problem is in the {\it thermodynamic limit}.  There
are no finite-size effects in infinite-dimensions.

Static two-particle properties are also easily calculated since the
irreducible vertex function is local~\cite{zlatic_horvatic_infd}.  The
static susceptibility for CDW order is given by
\begin{eqnarray}
\chi^{CDW}({\bf q})&\equiv& {1\over2N}\sum_{jk\sigma\sigma'}
e^{i{\bf q}\cdot
({\bf R}_j-{\bf R}_k)} T\int_{0}^{\beta} d\tau \int_{0}^{\beta} d\tau '\cr
& &\quad\quad\quad\quad\quad\quad\quad\quad\quad
[\langle n_{j\sigma}(\tau) n_{k\sigma'}(\tau ')\rangle -
\langle n_{j\sigma}(\tau)\rangle\langle n_{k\sigma'}
(\tau ')\rangle ] \cr
&\equiv& T\sum_{mn} \tilde\chi^{CDW} ({\bf q},i\omega_m,i\omega_n)=
T\sum_{mn} \tilde\chi_{mn}^{CDW} ({\bf q})\quad,
\label{eq: chicdw}
\end{eqnarray}
at each ordering wavevector ${\bf q}$.  Dyson's equation for the two-particle
Green's function becomes~\cite{jarrell,zlatic_horvatic_infd}
\begin{equation}
\tilde\chi_{mn}^{CDW}({\bf q})=\tilde\chi_{m}^0({\bf q})\delta_{mn}
-T\sum_p \tilde\chi_m^0({\bf q})\Gamma_{mp}^{CDW}\tilde
\chi_{pn}^{CDW}({\bf q})\quad ,
\label{eq: cdwdys}
\end{equation}
with $\Gamma_{mn}^{CDW}$ the (local) irreducible vertex function in the CDW
channel.

The
bare CDW susceptibility $\tilde\chi_n^0({\bf q})$ in
Eq.~(\ref{eq: cdwdys}) is defined in terms of the dressed
single-particle Green's function
\begin{eqnarray}
\tilde\chi_n^0({\bf q})&\equiv& -{1\over N} \sum_{\bf k} G_n({\bf
k})G_n({\bf k+q})
=-{1\over{\sqrt{\pi}}}{1\over{\sqrt{1-X^2}}}\cr
& &\quad\quad\times\int_{-\infty}^{\infty}
dy {{e^{-y^2}}\over{i\omega_n+\mu-\Sigma_n-y}}F_{\infty}\left [ {{i\omega_n+
\mu-\Sigma_n-Xy}\over{\sqrt{1-X^2}}}\right ]
\label{eq: chi0cdw}
\end{eqnarray}
and all of the wavevector dependence is included in the
scalar~\cite{brandt_mielsch,muellerhartmann} $X({\bf q})
\equiv \sum\nolimits_{j=1}^d \cos {\bf q}_j/d$.  The mapping ${\bf q}
\mapsto X({\bf q})$ is a many-to-one mapping that determines an equivalence
class of wavevectors in the Brillouin zone.  ``General'' wavevectors are
all mapped to $X=0$ since $\cos {\bf q}_j$ can be thought of as a random
number between $-1$ and 1 for ``general'' points in the Brillouin zone.
Furthermore, all possible values of $X$ $(-1\le X\le 1)$ can be labeled
by a wavevector that lies on the diagonal of the first Brillouin zone extending
from the zone center $(X=1)$ to the zone corner $(X=-1)$.  The irreducible
vertex function $\Gamma_{mn}^{CDW}$ is determined by inverting the Dyson
equation in Eq.~(\ref{eq: cdwdys}) for the {\it local} susceptibility
(which is determined by the
Monte Carlo techniques of the following section).  Once
the irreducible vertex function is found, then Eq.~(\ref{eq: cdwdys}) is
employed to calculate the momentum-dependent susceptibility.

A similar procedure is used to explore the superconductivity of the model.
Here, as in the Hubbard model, it is only necessary to look for
superconductivity with the same symmetry as the lattice (s-wave) since
other superconductivity with other symmetries do not have pairing
interactions~\cite{jarr_prusch}.  For the singlet {\it s}-wave SC channel,
the corresponding definitions are as follows:  The static susceptibility
in the superconducting channel is defined to be
\begin{eqnarray}
\chi^{SC}({\bf q})&\equiv& {1\over N}\sum_{jk}
e^{i{\bf q}\cdot
({\bf R}_j-{\bf R}_k)} T\int_{0}^{\beta} d\tau \int_{0}^{\beta} d\tau '
\langle c_{j\uparrow}(\tau)
c_{j\downarrow}(\tau)c_{k\downarrow}^{\dag}(\tau ')c_{k\uparrow}^{\dag}(\tau ')
\rangle  \cr
&\equiv& T\sum_{mn} \tilde\chi^{SC} ({\bf q},i\omega_m,i\omega_n)=
T\sum_{mn} \tilde\chi_{mn}^{SC} ({\bf q})\quad,
\label{eq: chisc}
\end{eqnarray}
for superconducting pairs that carry momentum ${\bf q}$;  Dyson's equation
becomes
\begin{equation}
\tilde\chi_{mn}^{SC}({\bf q})=\tilde\chi_{m}^0{'}({\bf q})\delta_{mn}
-T\sum_p \tilde\chi_m^0{'}({\bf q})\Gamma_{mp}^{SC}\tilde
\chi_{pn}^{SC}({\bf q})\quad ,
\label{eq: scdys}
\end{equation}
with $\Gamma_{mn}^{SC}$ the corresponding irreducible vertex function for the
SC channel; the bare pair-field susceptibility becomes
\begin{eqnarray}
\tilde\chi_n^0{'}({\bf q})&\equiv& {1\over N} \sum_{\bf k} G_n({\bf
k})G_{-n-1}({\bf -k+q})
={1\over{\sqrt{\pi}}}{1\over{\sqrt{1-X^2}}}\cr
& &\times\int_{-\infty}^{\infty}
dy {{e^{-y^2}}\over{i\omega_n+\mu-\Sigma_n-y}}F_{\infty}\left [ {{-i\omega_{n}+
\mu-\Sigma_{n}^*-Xy}\over{\sqrt{1-X^2}}}\right ]
\label{eq: chi0sc}
\end{eqnarray}
with the special value
$\tilde\chi_n^0{'}(X=1)=-{\rm Im}G_n/{\rm Im}(i\omega_{n}-\Sigma_{n})$ for the
SC pair that carries no net momentum; and finally the irreducible vertex
function is also
determined by inverting the Dyson equation in Eq.~(\ref{eq: scdys}) for the
{\it local} susceptibility.

\bigskip
\noindent{\bf 2. Monte Carlo Algorithm}
\bigskip

\noindent The dynamics of the impurity problem are identical to that of an
impurity embedded in a host metal described by the ``bare'' Green's function
$G_0$ \cite{schweitzer_czycholl,georges_kotliar,jarrell}.
Thus, given $G_0$, the impurity problem may be solved by using the quantum
Monte Carlo (QMC) algorithm of Hirsch and Fye~\cite{hirsch_fye} (an alternative
derivation of this algorithm
is presented in the appendix).  In the QMC the problem is cast into a
discrete path formalism in imaginary time, $\tau_l$, where
$\tau_l=l\Delta\tau$, $\Delta\tau=\beta/L$, and $L$ is the number of times
slices.  The values of $L$ used ranged from $40$ to $160$, with the largest
values of $L$ reserved for the largest values of $\beta$  because the time
required by the algorithm
increases like $L^3$.  Since the bare Green's function $G_0^{-1}$ in
Eq.~(\ref{eq: gdef}) is
not {\it a priori} known, the QMC algorithm must be iterated
to determine a self-consistent solution for the Green's
function of the infinite-dimensional lattice.  The procedure~\cite{jarrell}
is to begin with a bare Green's function $G_0^{-1}$,
use the QMC algorithm to determine the self energy $\Sigma$,
calculate the lattice Green's function from Eq.~(\ref{eq: gloc}), and
determine a
new bare Green's function from Eq.~(\ref{eq: gdef}).  This process is
iterated until convergence is reached (typically $7-9$
iterations).  At each step, the precision
(the total number of field-configurations generated) is increased.
In addition, results from high temperature runs are used to initialize
lower temperature runs.  These last two steps, are commonly used to to anneal
out the state with the lowest free energy.

The details of the (Hirsch-Fye~\cite{hirsch_fye}) impurity algorithm, as
modified
for the Holstein-Hubbard model, are reproduced in the appendix.
For the remainder of this section, we will discuss the
modifications necessary to apply this algorithm to the infinite-dimensional
limit.   The main difficulty  is that the Hirsch-Fye algorithm requires an
imaginary-time path integral technique which only produces data
for $G(\tau)$ at a {\em{discrete}} set of points in Euclidean time
$0<\tau<\beta$; whereas, the self-consistency step requires either
the Matsubara frequency Green's function or the corresponding
self energy.  This involves a numerical approximation of the integral  in
Eq.~(\ref{eq: greendef}).
Fourier transforming discretely sampled data presents some well known
difficulties~\cite{numrecipe}.  The principle difficulty is that
Nyquist's theorem tells us that above some frequency
$\omega_n=1/2\Delta\tau$,
unpredictable results are produced by conventional quadrature techniques.
Typically this problem is overcome by fitting the discrete data $G(\tau)$ with
a smooth cubic spline, and then performing the integral on the
splined data~\cite{numrecipe}.  Since the integral on the splined data may
be sampled on a much finer grid than the original data, this
process is referred to as over sampling.

	However, a problem still remains at high frequencies,
since the resulting $G(i\omega_n)$ goes quickly to zero for frequencies
above the Nyquist cutoff $1/2\Delta\tau$.  This presents a
difficulty since causality requires that
\begin{equation}
\lim_{\omega_n\to\infty} G(i\omega_n) \sim \frac{1}{i\omega_n}\, .
\end{equation}
In order to maintain causality~\cite{causal} of the Matsubara frequency Green's
functions, we condition the Fourier transform with a perturbation
theory result.  That is, we write
\begin{equation}
G(i\omega_n)=G_{pt}(i\omega_n)+
\int_0^\beta d\tau e^{i\omega_n\tau}\left(G(\tau)-G_{pt}(\tau)\right)\,.
\end{equation}
where $G_{pt}$ is a Green's function obtained from perturbation theory, and
the integral here is performed by the oversampling method described above.

	There are two obvious advantages to this approach.
First, the integral goes to zero for frequencies greater than the
Nyquist frequency $1/2\Delta\tau$, so that the resulting Green's function
has the same asymptotic behavior as the perturbation theory result,
and is thus causal.  Second, often, the perturbation theory result
is asymptotically exact ({\it i.~e.} results from a high temperature
expansion {\it etc}.),
and this then presents a way of appending exact QMC results at low frequency
with asymptotically exact perturbation theory results at high frequency.
The flow chart for the resulting algorithm is shown in Fig.~1.

\bigskip
\epsfxsize=4.25in
\epsfysize=2.5in
\epsffile{fig1.eps}
{\bf Figure 1.} {\it Flowchart for the $d=\infty$ algorithm.  The symbol
$\cal{F}$ denote that a Fourier transform is to be performed by
oversampling, and ${\cal{F}}^{-1}$ denotes its inverse.}
\bigskip

Once convergence of the algorithm in Fig.~1 is reached, the physical
properties of the system are calculated with the QMC.  A variety
of two-particle properties may be calculated in the
QMC approach since the irreducible vertex function is also local.  For
most quantities, this is straight-forward; however, the two-particle
Green's functions $\chi_{mn}^{loc}$ are difficult
to measure efficiently.  For example, consider the local opposite-spin
particle-particle propagator
\begin{eqnarray}
\chi_{nm}^{loc} &=& \int^{\beta}_0 d{\tau}_1  \cdots d{\tau}_4
e^{[i\omega_m({\tau}_2-{\tau}_1) -i\omega_n({\tau}_3-{\tau}_4)]}\cr
&&\quad\quad\quad\quad\quad\quad\quad\times\langle T_{\tau}
c_{\uparrow}(\tau_4)c_{\downarrow}(\tau_3)c^{\dag}_{\downarrow}(\tau_2)
c^{\dag}_{\uparrow}(\tau_1){\rangle}\,.
\end{eqnarray}
For a particular configuration of the Hubbard-Stratonovich fields,
the Fermions are noninteracting, thus the
expectation value indicated by the angle brackets above may be evaluated
in two steps.  First, using Wick's theorem, its value is tabulated for each
field configuration $\{s_l,x_l\}$. Second, using  Monte Carlo techniques these
configurations are averaged over. After the first step, the equation becomes
\begin{equation}
\chi_{nm}^{loc} = \left\langle \int^{\beta}_0 d{\tau}_1
\cdots d{\tau}_4
e^{[i\omega_m({\tau}_2-{\tau}_1) -i\omega_n({\tau}_3-{\tau}_4)]}
g_{\uparrow}(\tau_4,\tau_1)g_{\downarrow}(\tau_3,\tau_2)
\right\rangle_{m.c.}
\end{equation}
where the $m.c.$ subscript means that the Monte Carlo average is
still to be performed.

To measure this on the computer, the integrals must be approximated
by sums.  Since the Green's functions change discontinuously when the two
time arguments intersect, the best integral approximation
that can be used here is the trapezoidal approximation.
Using this, we will run into Green's functions with
both time arguments the same $g(j,j)$.
This is stored as $g(j^+,j)$ (i.e.\ it is assumed that the first time
argument is slightly greater than the second), but in the sums we
clearly want the equal time Green's function to be the average
$\{g(j^+,j)~+g(j,j^+)\}/2~=
g(j^+,j)-1/2$.  If we call {\bf g}, with $1/2$ subtracted from
its diagonal elements, ${\bf \gb}$, then
\begin{eqnarray}
\chi_{nm}^{loc}&=&
\left\langle
\left( \sum_{j,k} \dtau~e^{+i\pi j (2n+1)/L}
\gb_{\uparrow} (j,k)~\dtau~e^{-i\pi k (2m+1)/L}\right)\right.\\
& &\left.\left( \sum_{p,q} \dtau~e^{-i\pi p (2n+1)/L}
\gb_{\downarrow} (p,q)~\dtau~e^{+i\pi q (2m+1)/L}\right)
\right\rangle_{m.c.} \nonumber
\end{eqnarray}
This measurement may be performed efficiently if each term in
parenthesis is tabulated first and stored as a matrix, and then the
direct product of the two matrices taken as the estimate of
${\bf{\chi}}^{loc}$.
When done this way, the time required for this measurement scales
like $\sim L^3 $ rather than $\sim L^4$ as would result from a
straight-forward evaluation of the sums implicit in Eq.~(20).

Finally, the irreducible vertex function is determined by inverting the
relevant
local Dyson equation.  The momentum-dependent susceptibility may
then be calculated from Eq.~(\ref{eq: cdwdys}) or (\ref{eq: scdys}).

\bigskip
\noindent{\bf 3. Results}
\bigskip

\noindent For the results presented here, we chose an intermediate phonon
frequency $\Omega=0.5t^*$ (which is approximately one-eight of the effective
electronic bandwidth) for which there is a competition between CDW and
SC order.  As shown in Fig.~3, CDW order is favored near half filling (due to
Fermi surface nesting) and SC order is favored away from half filling.  As
shown in Fig.~2b, there is a maximum CDW transition temperature, because it
decreases as the coupling strength increases in the strong-coupling regime.


\bigskip
\epsfxsize=4.5in
\epsffile{fig2.eps}
{\bf Figure 2}: {\it (a) Effective potential for the phonon (after integrating
out the effects of the electrons) as a function of electron-phonon coupling
and (b) the CDW transition temperature at half filling as a function of the
coupling.  The parameters chosen here are $\Omega=0.5t^*$ and $U_c=0$.  Note
that the maximum in the $T_c$ curve occurs when the barrier height of the
double-well potential is equal in magnitude to $T_c$.}
\bigskip
\vfill
\eject

In order to shed some light on the transition from weak to strong coupling
the QMC simulations were sampled to determine a time-averaged
effective phonon potential.  The probability $P(x)$ that the phonon
coordinate $x(\tau_{\ell})$ lies in the interval from $x$ to $x+\delta x$ was
calculated for each time slice $\tau_{\ell}$ and averaged over all time
slices.  An effective phonon potential $V_{eff.}(x)$ was then
extracted from the probability distribution
$P(x)\propto \exp [-\beta V_{eff.}(x)]$ \cite{schuttler}.
This effective potential is plotted in Fig. 2~(a) for four different values
of the electron-phonon coupling strength at a temperature $T=1/7$.
In the case of weak coupling $(g=0.325)$, the potential appears
harmonic.  The potential flattens when $U\approx t^*$ ($g=0.5$) and as $g$
increases further, a double-well structure develops \cite{yu_anderson}.  The
barrier height grows linearly
with $g$ as does the separation of the minima.  The peak of the $T_c(g)$
curve for the CDW transition [see Fig. 2(b)] is reached at the point
where the barrier height is on the order of $T_c$ $(g=0.625)$.  Beyond this
point $(g=1.0)$ the system enters the strong-coupling regime and $T_c$
decreases.

In the region where the double-well potential has developed, the phonon
coordinate tunnels between the wells and the tunneling rate decreases
as the temperature is lowered below the barrier height.  At this point
the system may be considered to be a random mixture of empty
sites and bipolarons that fluctuates in time.  Tunneling through the barrier
produces correlations between the empty-sites and the bipolarons resulting
in a condensed CDW phase.  However as the barrier height increases, the
transition
temperature drops because the tunneling is suppressed.  The transition
temperature reaches its maximum at the point where the barrier height
is equal in magnitude to $T_c$.

As the system is doped away from half-filling there is a
competition between CDW order
and superconductivity .  We find that the CDW-ordered state remains ``locked''
at the
``antiferromagnetic'' point ($X=-1$) for a wide range of dopings away from
half-filling.
Figure 3 displays the results for the transition temperature of the
Holstein-Hubbard model as a function of electron concentration
for two different values of $U_c$ at $g=0.5t^*$. In the case where $U_c=0$
the system must be doped out to a concentration of $\rho_e=0.52$ before
it becomes superconducting.  There was no evidence for any incommensurate
order when $U_c=0$.  We expect that a Coulomb repulsion will favor
the SC phase over the CDW phase because the Coulomb repulsion directly reduces
the CDW interaction, but is not as effective at reducing the SC interaction
because of the retardation between the pairing electrons which allows the
electrons to attract each other without being at the same site at the same time
(the so-called pseudopotential effect).  This result is clearly seen in
Figure~3, where the phase space for the SC order increases when $U_c=0.5$.
Note that a finite Coulomb repulsion also favors the appearance of
incommensurate phases, which now can be detected with the QMC techniques.

\bigskip
\epsfxsize=4.5in
\epsffile{fig3.eps}
{\bf Figure 3}: {\it Transition temperature for the Holstein-Hubbard model
at $\Omega=0.5t^*$ and $g=0.5t^*$.  As the Coulomb repulsion is increased,
the SC phase becomes more stable, as do incommensurate CDW phases.}
\bigskip

\bigskip
\noindent{\bf Acknowledgments}
\bigskip

\noindent We would like to acknowledge useful conversations with
H.\ Akhlaghpour,
D.~L.\ Cox,
R.~M.\ Fye,
Th.\ Pruschke,
R.\ Scalettar, and D.\ J.\ Scalapino.
This work was supported by the National Science Foundation
grant number DMR-9107563 and by the Office of Naval Research under Grant No.
N00014-93-1-0495.  In addition MJ would like to acknowledge the
support of the NSF NYI program. Computer support was provided by the Ohio
Supercomputer Center.

\bigskip
\noindent {\bf Appendix. Derivation of Quantum MC algorithm of Hirsch and Fye}
\bigskip

\noindent The purpose of this section is to derive the Hirsch-Fye
algorithm~\cite{hirsch_fye} using Grassmann algebra.  We begin
by splitting the single impurity Anderson model Hamiltonian, into bare
and interacting parts, $H = H_{0} + H_{1} + H_{2}$, where
\begin{eqnarray}
H_{0}&=&\sum_{k\si} \ep({\bf k}) c_{k\si}^{\dag} c_{k\si}
+ V \sum_{k\si} ( f_{\si}^{\dag} c_{k\si} + c_{k\si}^{\dag} f_{\si} )\cr
&+&\lep  gx-\mu \rip( n_{f\up}+n_{f\dwn}-1 ) +\frac12 M\Omega^2x^2\quad ,
\label{eq: h0}
\end{eqnarray}
\begin{equation}
H_{1} = U (n_{f\up}-1/2)(n_{f\dwn}-1/2)\,,
\label{eq: h1}
\end{equation}
and
\begin{equation}
H_{2} = \frac{p^2}{2M}\,.
\label{eq: h2}
\end{equation}
To obtain the Trotter-Suzuki decomposition for the partition
function~\cite{suzuki} we divide the interval $[0,\be]$
into $L$ sufficiently small subintervals such that the commutators
$\Delta\tau^2\left[H_0,H_I\right]$ {\it etc}.\ may be neglected. This leads to
\begin{equation}
Z=Tre^{-\be H}=Tr\prod_{l=1}^{L}e^{-\De\tau H}
	\approx Tr\prod_{l=1}^{L}e^{-\De\tau H_{0}}e^{-\De\tau H_{1}}
e^{-\De\tau H_{2}}\,.
\label{eq: trotter}
\end{equation}
The interacting part of the Hamiltonian, $H_{1}$, may be further decoupled
by mapping it to an auxiliary Ising field via a discrete
Hirsch-Hubbard-Stratonovich~\cite{HHS} transformation,
\begin{equation}
e^{-\De\tau H_{I}} = e^{-\De\tau U(n_{f\up}-1/2)(n_{f\dwn}-1/2) }
  = {1\over 2} e^{-\De\tau U/4} \sum_{s=\pm1}e^{\al s (n_{f\up} - n_{f\dwn})}
\label{eq: hhs}
\end{equation}
where $\cosh(\al)=e^{\De\tau U/2}$.
Finally, one may cast Eq.~(\ref{eq: trotter}) into functional-integral
form by using
coherent states [Grassmann variables for Fermions, and complex numbers
for the Bosons, $a=\sqrt{m\Omega/2}(x+ip/m\Omega)$ and $a^*$]. If we
integrate out the host Fermionic degrees of freedom $\{c_{k\si}\}$
as well as the momentum of the phonon, then we end up  with
\begin{equation}
S_{eff}=\left(\De\tau
V\right)^2\sum_{l,l',\si}f_{\si,l}^{*}G_{0}(l,l')f_{\si,l'} +
S_{int} + S_B\,,
\label{eq: seffapp}
\end{equation}
where
\begin{equation}
S_B=\frac{\Delta\tau}{2} \sum_{l} \left [ {\lep \frac{x_l-x_{l+1}}
{\Delta\tau}\rip^2 } \right ]
+\Omega^2 x_l^2\,,
\label{eq: sb}
\end{equation}
\begin{eqnarray}
S_{int}&=&\sum_{l,\si,x_l}-f_{\si,l}^{*}(f_{\si,l}-f_{\si,l-1}) \ \cr
       &+&      \De\tau f_{\si,l}^{*}
		( gx_l-\mu+{U\over 2}
                +\frac{\al}{\De\tau}s_{l}\si)f_{\si,l-1}\quad ,
\label{eq: sint}
\end{eqnarray}
and
\begin{equation}
G_{0}^{-1}(l,l') = \frac{1}{N} \sum_{k}\de_{l,l'} - \de_{l-1,l'}[ 1 - \De\tau
\ep({\bf k}) ]\ .
\label{eq: g0inv}
\end{equation}
At this point the correspondence of the impurity and the infinite-dimensional
Hubbard model is clear.  In both, $G_0$ contains the information about the
host into which the impurity is embedded.  The difference is that $G_0$ must
be determined self-consistently for the lattice model.

We will now proceed to derive the Monte Carlo algorithm~\cite{hirsch_fye}
sufficient for either the impurity or the infinite-dimensional lattice
problem. By integrating over $f_{\si,l}$ we can write down the partition
function (neglecting a numerical prefactor), as
\begin{equation}
Z = \sum_{\{s_{l},x_l\}} det(G_{\up\{s_{l},x_l\}}^{-1})
det(G_{\dwn\{s_{l},x_l\}}^{-1})e^{-S_B}
\label{eq: partfun}
\end{equation}
where
\begin{equation}
G_{\si,\{s_{l},x_l\}}^{-1}(l,l') = \de_{l,l'} - \de_{l-1,l'}[ 1 -
\De\tau \lep gx_l-\mu\rip+ \al s_{l}\si ] - \De\tau^{2}V^{2} G_{0}
(l,l')
\label{eq: ginv}
\end{equation}
and we sum over all configurations of Hubbard-Stratonovich and phonon fields
$\{s_{l},x_l\}$. If we reexponentiate the above formula by defining
${\V}_{\si,\{s_{l},x_l\}}(l) \equiv \De\tau (gx_l -\mu +
\al s_{l}\si/\De\tau  )$, we can write it
in a simple notation as
\begin{equation}
G_{\si}^{-1} = 1 + T e^{{\V}_{\si}} + \De\tau^{2}V^{2}G_{0}\;,
\label{eq: ginv2}
\end{equation}
where T is $\de_{l-1,l'}$ and ${\V}_{\si} \equiv {\V}_{\si,\{s_{l},x_l\}}(l)$
for one special configuration. For another field configuration
the only difference comes from ${\V}_{\si}$ such that
${G'}_{\si}^{-1}-G_{\si}^{-1} = T (e^{{\V}'_{\si}}-e^{{\V}_{\si}}) +
O(\De\tau^{3/2})$ (note that $\al$ is of the order of $\De\tau^{1/2}$). On the
other hand
$T = (G_{\si}^{-1} -1 -\De\tau^{2}V^{2}G_{0})e^{-{\V}_{\si}}$ which results in
\begin{equation}
	 {G'}_{\si}^{-1}-G_{\si}^{-1} =
(G_{\si}^{-1}-1)e^{-{\V}_{\si}}(e^{{\V}'_{\si}}-e^{{\V}_{\si}}) +
O(\De\tau^{3/2})\,.
\label{eq: ginv3}
\end{equation}
Multiplying from the left by $G$ and from the right by ${G'}$ and, ignoring
terms $O(\De\tau^{3/2})$, we find
\begin{equation}
{G'}_{\si}=G_{\si}+(G_{\si}-1)(e^{{\V}'_{\si}-{\V}_{\si}}-1){G'}_{\si}\ ,
\label{eq: ginv4}
\end{equation}
or
\begin{equation}
G_{\si}{G'}_{\si}^{-1}=1+(1-G_{\si})(e^{{\V}'_{\si}-{\V}_{\si}}-1)\quad .
\label{eq: ghinv5}
\end{equation}
The probability of having a configuration $\{s_{l},x_l\}$ is
$P_{sx} \propto det(G_{\up\{s_{l},x_l\}}^{-1})$
\linebreak
$\times det(G_{\dwn\{s_{l},x_l\}}^{-1})e^{-S_B}$;
on the other hand the detailed balance requires
\begin{equation}
P_{sx'}P_{sx'\rightarrow sx} = P_{sx}P_{sx\rightarrow sx'}\quad ,
\label{eq: detbal}
\end{equation}
for all $sx'$.
We may satisfy this requirement by defining the probability of going from
${\{s_{l},x_l\}}$ to ${\{s'_{l},x'_l\}}$ as $R/(1+R)$,
where
\begin{equation}
R\equiv \frac{ det({G'}_{\up})det({G'}_{\dwn}) e^{-S'_B}}{
det(G_{\up})det(G_{\dwn}) e^{-S_B}}
\label{eq: rdef}
\end{equation}
is the relative weight of two configurations.
If the difference between two configuration is due to a flip of
a single Hirsch-Hubbard-Stratonovich field at the $m$th imaginary time
slice then~\cite{hirsch_fye}
\begin{equation}
R = \prod_{\si}[1 + (1-G_{\si m,m})(e^{-2\al\si s_{m}} -1)]  \,,
\label{eq: rnew}
\end{equation}
or, if the difference is due to a change in the phonon displacement
$x_l\to x'_l$, then
\begin{equation}
R = e^{S'_B-S_B}
\prod_{\si}[1 + (1-G_{\si m,m})(e^{\Delta\tau g (x_l-x'_l)} -1)] \,.
\label{eq: rnew2}
\end{equation}

Finally we can write down the evolution of the Green's function in
the QMC time, when for example we flip a single Hirsch-Hubbard-Stratonovich
field at the $m$th imaginary time slice~\cite{hirsch_fye}
\begin{eqnarray}
{G'}_{\si i,j}&=&G_{\si i,j}+(G_{\si i,m}-\de_{i,m})
	(e^{-2\al\si s_{m}}-1)\cr
       & & \quad\quad\quad\times \{1+(1-G_{\si m,m})
	(e^{-2\al\si s_{m}}-1)\}^{-1}G_{\si m,j}\,.
\label{eq: gupdate}
\end{eqnarray}
Similarly, when we change a single boson $x_m$ at the $m$th time slice,
\begin{eqnarray}
{G'}_{\si i,j}&=&G_{\si i,j}+(G_{\si i,m}-\de_{i,m})
	(e^{\Delta\tau g (x_l-x'_l)}-1)\cr
       & & \quad\quad\quad\times \{1+(1-G_{\si m,m})
	(e^{\Delta\tau g (x_l-x'_l)}-1)\}^{-1}G_{\si m,j}\,.
\label{eq: gupdate2}
\end{eqnarray}

	The QMC process precedes by sequentially proposing changes in
each field, accepting these changes with probability $P_{sx->sx'}$, and
updating the Green's function with Eq.~(\ref{eq: gupdate}) or
Eq.~(\ref{eq: gupdate2}) when the change is accepted.  In addition to using
local moves in which a single spin or a single phonon field is changed,
we also employ global
moves, in which either all of the spins are flipped, or all of the
phonon coordinates are shifted,


\begin{thebibliography}{99}
\bibitem{causal} Obeying causality is crucial to guarantee that the
procedure described here is properly Herglotz.
\end{thebibliography}
\end{document}